\documentclass[a4paper,11pt]{scrartcl}

\usepackage[english]{babel}
\usepackage[utf8]{inputenc}
\usepackage[hyphens]{url}
\usepackage{hyperref}
\usepackage{graphicx}
\usepackage{amsmath}
\usepackage{amssymb}
\usepackage{subcaption}
\usepackage{xspace}
\usepackage{booktabs}
\usepackage{lipsum}
\usepackage{enumitem}
\usepackage{float}
\usepackage{xcolor}
\usepackage{todonotes}
\usepackage{MnSymbol}
\usepackage{numprint}
\usepackage{csquotes}
\usepackage{multirow}

\usepackage[eprint=false]{biblatex}
\renewbibmacro*{doi+eprint+url}{%
	\printfield{doi}%
	\newunit\newblock%
	\iftoggle{bbx:eprint}{%
		\usebibmacro{eprint}%
	}{}%
	\newunit\newblock%
	\iffieldundef{doi}{%
		\usebibmacro{url+urldate}}%
	{}%
}
\bibliography{ms}

\usepackage{microtype}

\usepackage{listings}

\usepackage{textcomp}
\definecolor{keyword}{HTML}{2771a3}
\definecolor{pattern}{HTML}{b53c2f}
\definecolor{string}{HTML}{be681c}
\definecolor{relation}{HTML}{7e4894}
\definecolor{variable}{HTML}{107762}
\definecolor{comment}{HTML}{8d9094}

\newcommand{\danger}{{\fontencoding{U}\fontfamily{futs}\selectfont\char 66\relax}}
\newcommand{\ordo}[1]{\mathcal{O}\left( #1 \right)}

\newcommand{\ie}{i.e.\@\xspace}

\newcommand{\eg}{e.g.\@\xspace}

\newcommand{\yes}{$\bigotimes$\xspace}
\newcommand{\maybe}{$\bigoslash$\xspace}
\newcommand{\no}{$\bigcircle$\xspace}

\newcommand{\yedscale}{0.56}

\begin{document}

\title{An analysis of the SIGMOD 2014 Programming Contest: Complex queries on the LDBC social network graph}
\author{Márton Elekes\footnote{Budapest University of Technology and Economics, Department of Measurement and Information Systems, Hungary}
    \and
    János Benjamin Antal
    \and
    Gábor Szárnyas\footnote{CWI, Amsterdam, The Netherlands. Corresponding author. Email: \url{szarnyasg@gmail.com}}}
\date{}
\maketitle

\begin{abstract}
This report contains an analysis of the queries defined in the SIGMOD 2014 Programming Contest.
We first describe the data set, then present the queries, providing graphical illustrations for them and pointing out their caveats.
Our intention is to document our lessons learnt and simplify the work of those who will attempt to create a solution to this contest.
We also demonstrate the influence of this contest by listing followup works which used these queries as inspiration to design better algorithms or to define interesting graph queries.
\end{abstract}

\section{Introduction}

The SIGMOD conference hosts an annual Programming Contest where teams of graduate students are required to solve database-related programming tasks.
Teams have approximately 3 months to implement their solution and compete on a previously agreed metric (\eg lowest execution time).
The Programming Contest of SIGMOD 2014%
\footnote{\url{https://web.archive.org/web/20210118224923/www.cs.albany.edu/~sigmod14contest/}}
focused on graph processing problems and consisted of 4 complex graph queries on the LDBC Social Network Benchmark's schema~\cite{Angles2020}.
The queries contained a mix of relational operators (such as filtering and aggregation) along with graph analytical computations (such as breadth-first search and connected components).
This document contains an overview and analysis of the data set and queries of the 2014 Programming Contest.

\section{Graph schema and data sets}

\begin{figure}[htb]
    \centering
    \includegraphics[scale=\yedscale]{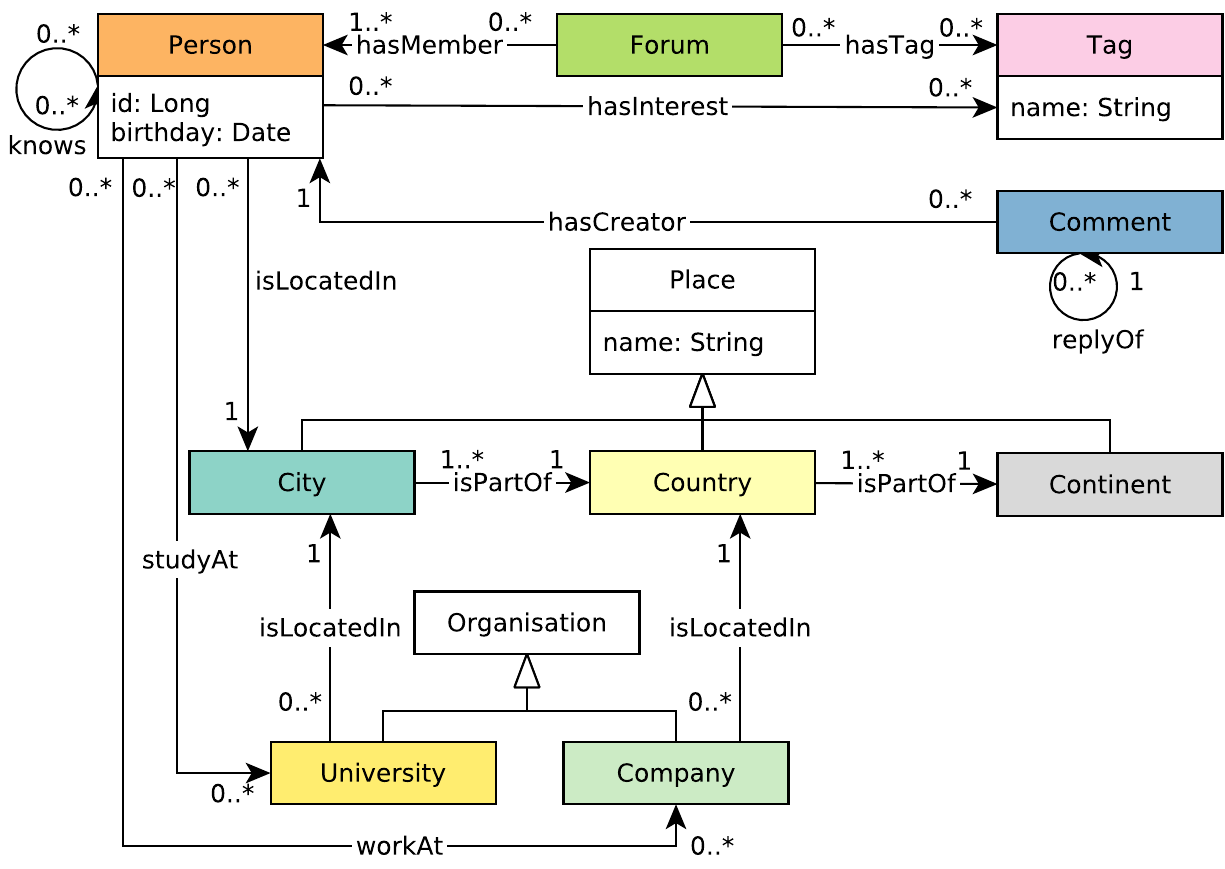}
    \caption{Schema of the social network graphs. Only the relevant node/edge types and property names (along with their types) shown.}
    \label{fig:schema}
\end{figure}

\subsection{Graph schema}
The social network graph instances used in the contest can be represented as a \emph{property graph}~\cite{DBLP:journals/csur/AnglesABHRV17} that conforms to the schema of the LDBC Social Network Benchmark.
The relevant part of the schema is shown in \autoref{fig:schema} (the full schema can be found in~\cite{Angles2020}).
The edges in the graph are directed with the exception of the \texttt{knows} edges which are treated as undirected.

\subsection{Data sets}
The data sets containing the graph instances are produced by the LDBC Datagen~\cite{DBLP:conf/tpctc/PhamBE12}, a Hadoop-based graph generator.
Datagen generates a realistic Facebook-like power-law degree distribution for the \textsf{Person}--\texttt{knows}--\textsf{Person} graph and also enforces certain correlations, \eg people who studied together are more likely to become friends.

The data sets used in the contest are available in the data repository~\cite{cwi:ldbc-sigmod-data-sets} at \url{https://repository.surfsara.nl/datasets/cwi/ldbc-sigmod-data-sets}.
This includes the contest's original data sets sets with 1k and 10k \textsf{Persons} (\texttt{o1k}, \texttt{o10k}).
To generate larger data sets, use the LDBC Datagen tagged as \texttt{sigmod2014contest}%
\footnote{\url{https://github.com/ldbc/ldbc_snb_datagen/releases/tag/sigmod2014contest}}
with the configurations provided in \autoref{sec:datagen-params}.
Data sets with 1k, 10k, 100k, 1M persons are also available in the data repository (\texttt{p1k}, \texttt{p10k}, \texttt{p100k}, \texttt{p1000k}).

\paragraph{Caveat \danger}
In the generated data, \textsf{Places} at different hierarchy levels, \ie \textsf{Continents}, \textsf{Countries}, and \textsf{Cities} can have the same name.
For example, ``Australia'' is both a \textsf{Continent} and a \textsf{Country}, while ``Indonesia'' is both a \textsf{Country} and a \textsf{City}.%
\footnote{The Datagen version used in the contest generates multiple erroneous \textsf{City} names such as ``India'' and ``Indonesia''.
    These have been removed in later versions of Datagen.
    However, even in the latest version of the data generator, \textsf{Place} names are not completely disjoint at different hierarchy levels as ``Australia'' is (correctly) both generated as a \textsf{Country} and a \textsf{Continent}.}

\begin{lstlisting}[language=bash]
$ grep Australia place.csv
1463|Australia|http://dbpedia.org/resource/Australia|Continent
4|Australia|http://dbpedia.org/resource/Australia|Country
    
$ grep Indonesia place.csv
39|Indonesia|http://dbpedia.org/resource/Indonesia|Country
661|Indonesia|http://dbpedia.org/resource/Indonesia|City   
\end{lstlisting}

\section{Queries}

The contest defines 4 queries.
In the contest, solutions were expected to execute a mix of these queries (see Section \ref{sec:contest-setup}).
In the following, we present the queries including their textual specification, their input parameters, output attributes and the illustration of their graph pattern.

\begin{enumerate}[label=\textbf{Q\arabic*.}]

    \begin{figure}[htb]
        \centering
        \includegraphics[scale=\yedscale]{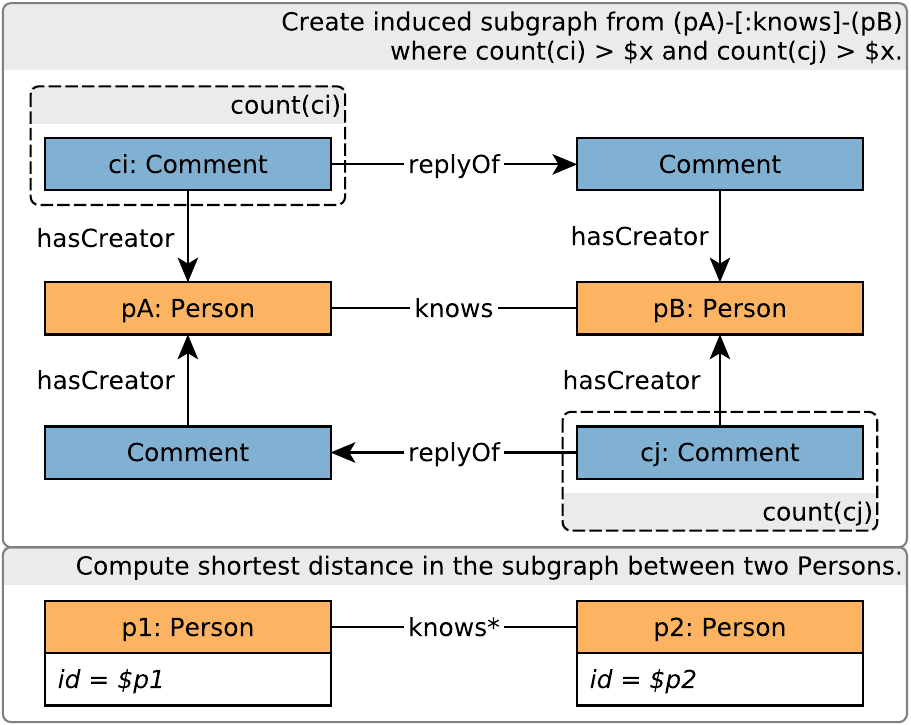}
        \caption{Graph pattern of Q1.}
        \label{fig:query1}
    \end{figure}

    \item \textbf{Shortest Distance over Frequent Communication Paths} (\autoref{fig:query1}).
          Given two integer \textsf{Person} identifiers $p1$ and $p2$, and another integer $x$, find the minimum number of hops (\ie the shortest distance) between $p1$ and $p2$ in the graph induced by \textsf{Persons} who

          \begin{itemize}
              \item know each other and
              \item communicate frequently with each other, \ie both have made more than $x$ \textsf{Comments} in reply to the other one's \textsf{Comments}.
          \end{itemize}

          \paragraph{Caveat \danger} Some remarks regarding Q1:

          \begin{itemize}
              \item The frequent communication has to happen \emph{both ways} between \textsf{Person} pairs.
              \item When determining the shortest distance, only the edges of the induced subgraph can be used.
              \item Notice that for $x = -1$, the original \textsf{Person}--\texttt{knows}--\textsf{Person} is equivalent to the induced subgraph, so it is not necessary to compute the number of interactions.
          \end{itemize}

          \begin{description}
              \item[API] \texttt{query1(p1, p2, x)}
              \item[Output] One integer (the hop count) per line.
              \item[Samples] \texttt{1k-sample-queries1.txt} and \texttt{1k-sample-answers1.txt} (\autoref{sec:sample-queries-and-answers})
          \end{description}

          \begin{figure}[htb]
              \centering
              \includegraphics[scale=\yedscale]{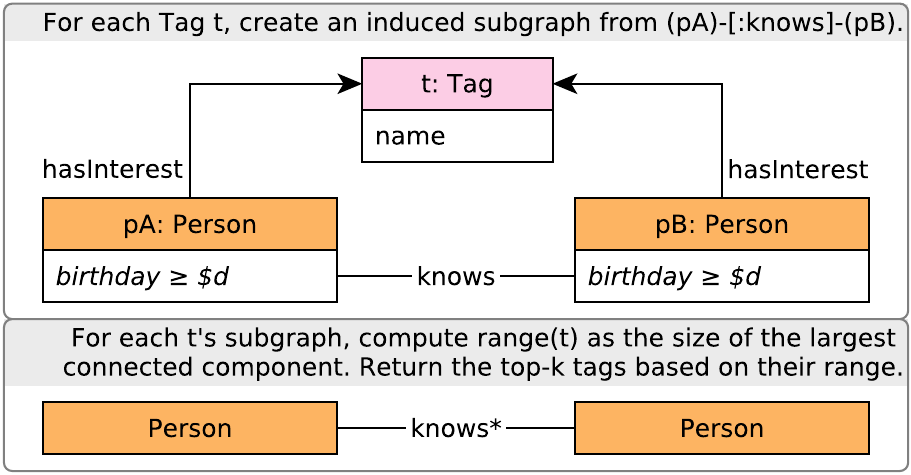}
              \caption{Graph pattern of Q2.}
              \label{fig:query2}
          \end{figure}

    \item \textbf{Interests with Large Communities} (\autoref{fig:query2}).
          Given an integer $k$ and a birthday $d$, find the top-$k$ \textsf{Tags}. A \textsf{Tag} is characterized with its \emph{range}, \ie the size of the largest connected component in the graph induced by \textsf{Persons} who
          \begin{itemize}
              \item know each other,
              \item are interested in the \textsf{Tag}, and
              \item were born on day $d$ or later.
          \end{itemize}

          \paragraph{Caveat \danger}
          When determining the connected components, only the \texttt{knows} edges in the induced subgraph should be used.

          \begin{description}
              \item[API] \texttt{query2(k, d)}
              \item[Output] Exactly $k$ strings (separated by a space) per line. These $k$ strings represent \textsf{Tag} names of interest, ordered by range from largest to smallest, with ties broken by lexicographical ordering, ascending (\eg ``A'' precedes ``B'' in the results).
              \item[Samples] \texttt{1k-sample-queries2.txt} and \texttt{1k-sample-answers2.txt} (\autoref{sec:sample-queries-and-answers})
          \end{description}

          \begin{figure}[htb]
              \centering
              \includegraphics[scale=\yedscale]{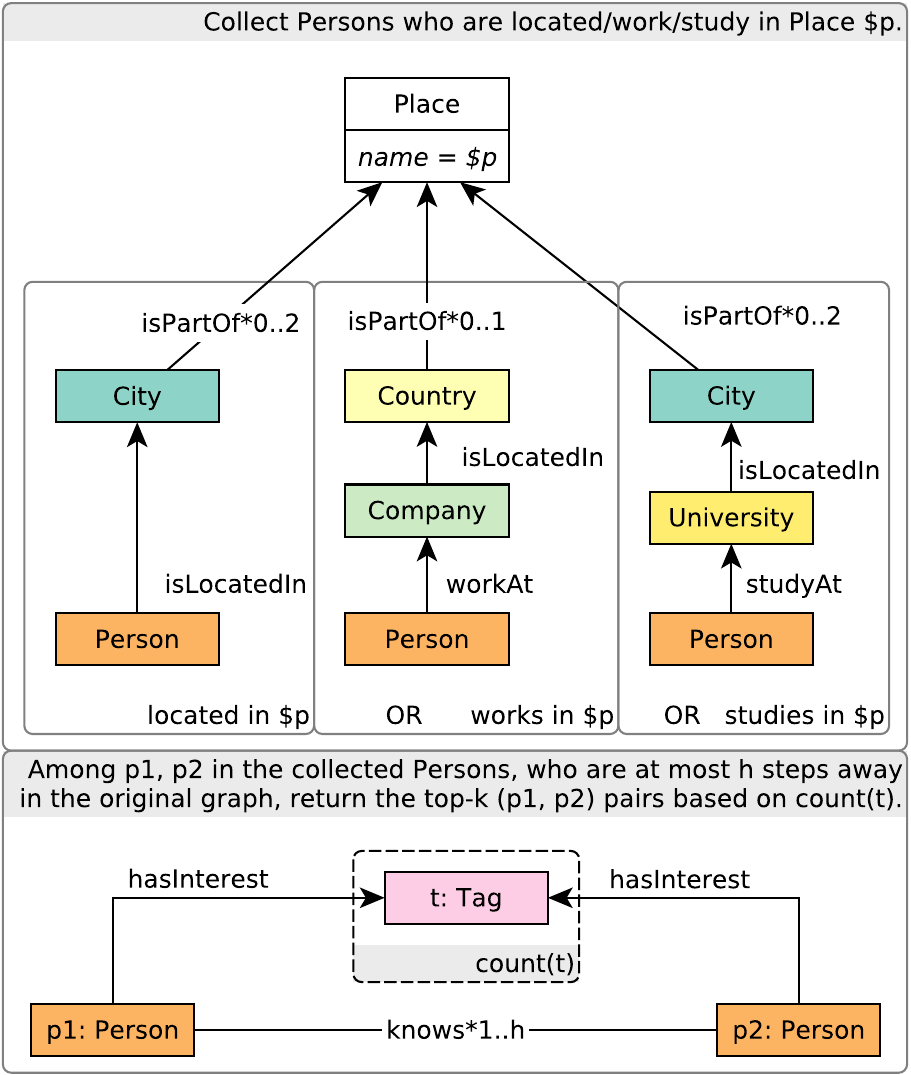}
              \caption{Graph pattern of Q3.}
              \label{fig:query3}
          \end{figure}

    \item \textbf{Socialization Suggestion} (\autoref{fig:query3}).
          Given an integer $k$, an integer maximum hop count $h$, and a \textsf{Place} name $p$, find the top-$k$ similar pairs of \textsf{Persons} based on the number of common interest \textsf{Tags}. For each of the $k$ pairs mentioned above, the two \textsf{Persons} must be located in $p$ or study/work at \textsf{Organisations} in $p$. Furthermore, these two \textsf{Persons} must be no more than $h$ hops away from each other in the original \textsf{Person}--\texttt{knows}--\textsf{Person} graph (i.e. their shortest distance is at most $h$).

          \paragraph{Caveat \danger}
          When determining the \textsf{Person} pairs with at most $h$ steps away, only the \textsf{Person} nodes in the selected \textsf{Place} are considered \emph{for selecting the pairs} but all \textsf{Person} nodes and \texttt{knows} edges should be considered \emph{when determining the shortest distance of a given pair}.

          \begin{description}
              \item[API] \texttt{query3(k, h, p)}
              \item[Output] Exactly $k$ pairs of \textsf{Person} ids per line. These pairs are separated by a space and \textsf{Person} ids are separated by the pipe character ``\texttt{|}''. For any \textsf{Person} id \texttt{p}, \texttt{p|p} must be excluded. For any pairs \texttt{p1|p2} and \texttt{p2|p1}, the second pair in lexicographical order must be excluded. These $k$ pairs must be ordered by similarity from highest to lowest, with ties broken by lexicographical ordering based on ascending numerical order of the ids.\footnote{Numerical ordering means the regular arithmetic ordering, \ie $9 < 10$. (A common issue is comparing numbers as strings, which would give the opposite order. This should be avoided here.)}
              \item[Samples] \texttt{1k-sample-queries3.txt} and \texttt{1k-sample-answers3.txt} (\autoref{sec:sample-queries-and-answers})
          \end{description}

          \begin{figure}[htb]
              \centering
              \includegraphics[scale=\yedscale]{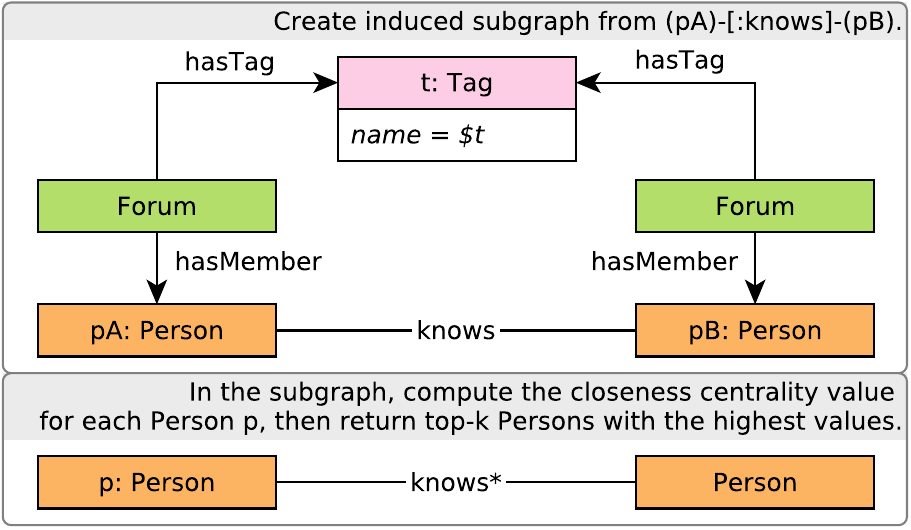}
              \caption{Graph pattern of Q4.}
              \label{fig:query4}
          \end{figure}

    \item \textbf{Most Central People} (\autoref{fig:query4}).
          Given an integer $k$ and a \textsf{Tag} name $t$, find the top-$k$ \textsf{Persons} based on the \emph{closeness centrality value} ($\mathit{CCV}$) in the graph induced by \textsf{Persons} who
          \begin{itemize}
              \item know each other and
              \item are members of \textsf{Forums} which have \textsf{Tag} $t$.
          \end{itemize}

          For each \textsf{Person} $p$ in the induced subgraph, compute
          $$\mathit{CCV}(p) =
              \frac%
              {\left( |C(p)| - 1 \right)^2}%
              {\left( n-1 \right) \cdot s(p)},
          $$
          where
          \begin{itemize}
              \item $C(p)$ denotes the nodes in the connected component of node $p$,
              \item $n$ is the number of nodes in the induced subgraph, and
              \item $s(p)$ is the sum of shortest distances to all other reachable \textsf{Persons} from $p$, \ie $$s(p) = \sum_{r \in C(p)} \mathrm{dist}(p,r) $$              
          \end{itemize}
          For isolated nodes, the divisor in the $\mathit{CCV}$ formula is $0$ due to $s(p) = 0$. For these, the centrality value is defined as $0$.

          \paragraph{Caveat \danger}
          When determining the $\mathit{CCV}$ value, only the \texttt{knows} edges in the induced subgraph should be used.
\end{enumerate}

\begin{description}
    \item[API] \texttt{query4(k, t)}
    \item[Output] Exactly $k$ \textsf{Person} ids (separated by a space) per line. These \textsf{Person} ids are ordered by centrality from highest to lowest, with ties broken by \textsf{Person} id, in ascending order (\ie id 1 precedes id 2 in the results).
    \item[Samples] \texttt{1k-sample-queries4.txt} and \texttt{1k-sample-answers4.txt} (\autoref{sec:sample-queries-and-answers})
\end{description}

\section{Analysis}

\subsection{Data set}

\paragraph{Distribution}

The power-law distribution of the \textsf{Person}--\texttt{knows}--\textsf{Person} graph implies that this graph has a small diameter. This means that techniques such as \emph{direction-optimizing traversal}~\cite{DBLP:conf/sc/BeamerAP12} (also known as \emph{push/pull} and \emph{top-down/bottom-up}) can be used to improve the performance of graph traversals.

\paragraph{Node id relabelling}

The nodes in the generated input graphs have \emph{sparse ids}, \ie their identifiers can take any 64-bit unsigned integer value.
These identifiers cannot be used directly by systems relying on compressed matrix representations such as CSR (Compressed Sparse Row)~\cite{DBLP:books/daglib/0009092}.
This necessitates the use of \emph{dense ids} with consecutive values in the $[0, |V|)$ range.
To provide such identifiers, systems need to perform \emph{node relabelling}, also known as \emph{dense vertex relabelling}~\cite{DBLP:conf/grades/ThenKK016}, \emph{vertex permutation}~\cite{DBLP:journals/corr/abs-1806-01799}, and \emph{mapping from sparse to dense keys}~\cite{DBLP:journals/pvldb/MuhlbauerRSRK013}.
In the context of this contest, implementers have to consider the tradeoffs of this approach.
On the one hand, performing such a relabelling is an expensive step.
On the other hand, it can improve the locality of neighbourhood lookup operations (compared to \eg hash-based lookups) which form a significiant portion of the operations in the queries.

\subsection{Queries}

We present a brief discussion on the complexity of implementing the queries of the contest efficiently.
The key algorithms (kernels) required by each query are listed in \autoref{tab:algorithms}.

\begin{table}
    \centering
    \begin{tabular}{@{}llcccc@{}}
        \toprule
        \multicolumn{1}{c}{Algorithm}                 & \multicolumn{1}{c}{Variant}  & Q1   & Q2   & Q3   & Q4     \\ \midrule
        \multirow{3}{*}{Unweighted shortest distance} & Single-source, single-target & \yes & \no  & \no  & \no    \\
                                                      & Multi-source                 & \no  & \no  & \yes & \no    \\
                                                      & All-pairs                    & \no  & \no  & \no  & \yes   \\ \midrule
        Weakly connected components                   & --                           & \no  & \yes & \no  & \maybe \\
        \bottomrule
    \end{tabular}
    \caption{Algorithms to be used for efficient evaluation of the queries.
        \emph{Notation} --
        \yes:~required,
        \no:~not required,
        \maybe:~applicable but not required.}
    \label{tab:algorithms}
\end{table}

\begin{enumerate}[label=\textbf{Q\arabic*.}]
    \item \textbf{Shortest Distance over Frequent Communication Paths}

          This query computes the (unweighted) shortest distance between two fixed nodes in an induced subgraph.
          The second part of the query, which determines the shortest distance, is rather straightforward: it can be implemented using a bidirectional search on the subgraph.

          The difficulty of this query lies in amalgamating its first part, which computes the induced subgraph, with its second part.
          For the majority of input parameters $x \geq 0$, \ie the induced subgraph is not equivalent to the full \textsf{Person}--\texttt{knows}--\textsf{Person} graph. In these cases, there are two key approaches for implementing the subgraph computation:

          \begin{description}
              \item[Approach~(1):] Precompute the induced subgraph and run the search on that graph.

              \item[Approach~(2):] For each edge, check the number of its interactions on-the-fly during the search.
          \end{description}

          Approach (1) is simpler to implement but often leads to suboptimal performance. This is especially the case if the path exists and is found quickly without having to traverse the entire subgraph, leaving most of the induced subgraph unused.

          Therefore, approach (2) can often avoid a lot of unnecessary computation. However, implementing this approach efficiently is complicated by the setup of the original contest, where solutions were expected to evaluate multiple queries at the same time.
          In particular, when evaluating multiple queries, computing the number of interactions for each edge on-the-fly might result in redundant computations as different traversals will likely reach the same edges. To prevent redundant computations, implementations should use some (thread-safe) caching mechanism to store the number of interactions for each edge.

    \item \textbf{Interests with Large Communities}

          Unlike the other queries which use a single induced subgraph, this query defines \emph{multiple subgraphs}, one for each \textsf{Tag}.
          There are 1457 \textsf{Tags} in the 1k \textsf{Person} data set and more than \numprint{10000} \textsf{Tags} in the 1M \textsf{Person} data set.
          Due to this and due to the fact that the computations are completely disjoint, this query is easy to parallelize by processing different \textsf{Tags} on different threads.

          To allow efficient evaluation of the selection on the birthday attributes (on the inequality condition $\textsf{birthday} \geq d$), it is recommended to define an index on the \textsf{Person.birthday} attribute.

          Once an induced subgraph has been created, solutions need to run a connected components algorithm to determine the size of its largest connected component. There are three key approaches to compute these:

          \begin{itemize}
              \item A na\"ive solution is to run repeated BFS traversals, each discovering a connected component and then restarting the BFS from a yet unvisited node, until all nodes are visited. While this approach often results in small BFS operations which often discover components of just a few nodes, it can provide a quick solution with acceptable performance in the context of this query.

              \item Solution authors often opt to use \emph{Tarjan's strongly connected components algorithm}~\cite{DBLP:journals/siamcomp/Tarjan72}. While a sequential variant of this algorithm is simple to implement, it has a complexity of $\ordo{|V| + |E|}$ and it is difficult to parallelize.

              \item Authors can exploit that this query only needs \emph{weakly connected components}.
              This can be computed with algorithms such as
              the \emph{Shiloach-Vishkin algorithm}~\cite{DBLP:journals/jal/ShiloachV82} with a complexity of $\ordo{|E| \cdot \log |V|}$, and
              the recent \emph{Afforest algorithm}~\cite{DBLP:conf/ipps/0001BB18}, which has a worst-case complexity of $\ordo{|V| + |E|}$ but performs well for most practical data sets and is easy to parallelize.
          \end{itemize}

    \item \textbf{Socialization Suggestion}

          The first part of this query requires efficient lookups for \textsf{Persons} in a given \textsf{Place}. This means that they are
          located there and/or
          study at a \textsf{University} there and/or
          work at a \textsf{Company} there.
          These lookups can be assisted by representing the \textsf{Places} using \emph{nested intervals}~\cite{DBLP:journals/sigmod/Tropashko05}.

          The task in the second part of this query is to determine which pairs of \textsf{Persons} are reachable from each other using at most $h$ hops.
          This can be implemented as a \emph{multi-source BFS} starting from each \textsf{Person}, then advancing the frontier of the traversal for $h$ steps and maintaining the visited nodes for each source.
          Once we completed the traversal, for each source \textsf{Person}, the other \textsf{Persons} found in its visited nodes are the ones reachable within $h$ hops.

          A more efficient approach can exploit the symmetry of the undirected \textsf{knows edges} by running a \emph{multi-source bidirectional BFS}, which only advances the frontier for $\left\lceil \frac{h}{2} \right\rceil$ steps, then looks for intersections between the sets of visited nodes for each pair of traversals. (Odd $h$ values need to be treated as a special case by intersecting the set of visited nodes produced after $\left\lceil \frac{h}{2} \right\rceil - 1$ steps and $\left\lceil \frac{h}{2} \right\rceil$ steps.)

    \item \textbf{Most Central People}
    
          The first part of this query, \ie finding the \textsf{Persons} who are members of a \textsf{Forum} which has a given \textsf{Tag} is relatively simple.
          The difficulty lies in the second part, particularly in computing the $s(p)$ value for all \textsf{Person} nodes.
          This can be formulated as an \emph{all-pairs shortest distances} (also known as \emph{all-pairs unweighted shortest paths}) problem.
          While this problem is simpler than the \emph{all-pairs weighted shortest paths} problem (which can be tackled by \eg the Floyd-Warshall algorithm in $\ordo{n^3}$ time), it is still computationally intensive.

          For a node $p$, its $s(p)$ value can be computed simply by running a BFS traversal: at each step $l$, we increment $s(p)$ by $l \times \textit{the number of nodes found at level } l$.
          The challenge in this query is the sheer number of traversals to execute: on the 1M \textsf{Person} data set, for a popular \textsf{Tag} with many \textsf{Forums}, there might be more than \numprint{100000} member \textsf{Persons} whose closeness centrality values need to be computed.
          Therefore, to be competitive on this query, solutions need to employ some sort of optimization such as compression and efficient search space pruning.
          
          In the following, we list a few potential optimization ideas.
          Some of these are mutually incompatible (\eg it is not possible to use integer matrices and bitwise compression) but others can be combined to get an efficient implementation.
          We also advise the reader to consult the presentations and posters of the top-ranking teams for more ideas.

          \paragraph{Multi-source BFS}
          The large number of BFS traversals, all of which perform the same operation during traveral (incrementing the $s(p)$ value until a fixed-point) lend themselves to bulk processing. Multi-source BFS operations can process multiple nodes in bulk and improve locality during the computation~\cite{DBLP:journals/pvldb/ThenKCHPK0V14,HPEC2020_GraphBLAS_SIGMOD_Contest}.

          \paragraph{Compression}
          Top-ranking solutions in the contest employed bitwise operations to optimize the performance of multi-source BFS traversals~\cite{DBLP:journals/pvldb/ThenKCHPK0V14,DBLP:conf/edbt/KaufmannTK017}. Using bitmaps encoded as \texttt{UINT64} values has multiple advantages: it saves memory and allows 64 traversals to be handled at once. Additionally, multiple \texttt{UINT64} values can be batch processed using SIMD operations.

          \paragraph{Integer matrix multiplication} 
          The problem can be elegantly expressed as matrix multiplications on integer matrices
          as demonstrated in the algorithm of~\cite{DBLP:journals/jcss/Seidel95}, which requires $\ordo{\log \text{diameter(G)}}$ dense matrix multiplications.
          Using non-na\"ive matrix multiplication algorithms, this can result in a lower complexity than that of the multi-source BFS-based algorithms which require $\ordo{\text{diameter(G)}}$ traversal steps.
          This approach can optimized further:
          (1)~In a distributed setup, reducing communication costs between parts of the matrix multiplication can result in a significant improvement of performance~\cite{DBLP:conf/ipps/SolomonikBD13}.
          (2)~Using Tiskin's algorithm~\cite{DBLP:conf/icalp/Tiskin01,DBLP:journals/corr/SolomonikH15}, one can formulate the problem using multiplications on a dense and a sparse matrix, making it possible to exploit the sparsity of the adjacency matrix.

          \paragraph{Splitting the graph to connected components}
          The definition of $s(p)$ states that for each node, only the other reachable \textsf{Person} nodes are considered. Therefore, solutions might opt to first run a (weakly) connected components algorithm on the subgraph, then compute the CCV values on the components separately. This allows establishing tighter bounds for search space pruning (see below).
          The limitation of this approach is that computing the connected components is fairly expensive and for popular \textsf{Tags}, the graph often consist of only a few components so the benefits of this computation are limited.

          \paragraph{Heuristics-based search space pruning}
          When determining the $s(p)$ values, many \textsf{Person} nodes $p$ can be discarded early if we can prove that their final $s(p)$ values will be larger than the top-$k$ $s(p)$ values in the same component (implying that their $\mathit{CCV}$ value will be lower).
          We can exploit that the computation of $s(p)$ for nodes with a high $\mathit{CCV}$ value finishes early as (by definition) these belong to a central node from which the other reachable nodes are just a few hops away.
          Therefore, we can use get their final $s(p)$ values early and use them to prune traversals whose $s(p)$ values will be excessively high.

          Team \emph{blxlrsmb} suggested a simple lower bound for $s(p)$ values. According to this, after step $l$, the following lower bound holds:
          $$s(p) \geq
          \sum_{r \in C(p) \wedge (\mathrm{dist}(p,r) \leq l)} \mathrm{dist}(p,r) +
          \sum_{r \in C(p) \wedge (\mathrm{dist}(p,r) > l)} (l+1), $$
          where the first term is equivalent to the value of $s(p)$ after $l$ steps and the second term can be computed with a simple multiplication:
          $$\left| \{ r \in C(p) \text{ and } r \text{ is unvisited}  \} \right| \cdot (l+1)$$
        
          \paragraph{Other algorithms}
          Paper~\cite{DBLP:journals/talg/Chan12} presents a family of sophisticated algorithms for the \emph{unweighted undirected all-pairs shortest paths problem} guaranteeing a complexity of $\ordo{|V| \cdot |E|}$.

          \paragraph{Top-k closeness centrality}
          Paper~\cite{DBLP:conf/icde/OlsenLH14} (co-authored by Jeong-Hyon Hwang, one of the contest's organizers) presents a top-$k$ closeness centrality value algorithm.
\end{enumerate}

\subsection{Choke points}

\begin{table}
    \centering
    \begin{tabular}{@{}lcccc@{}}
        \toprule
        \multicolumn{1}{c}{Choke point}                   & Q1   & Q2   & Q3   & Q4   \\
        \midrule
        CP-1.3 Top-k pushdown                             & \no  & \yes & \yes & \yes \\
        CP-2.3 Join type selection                        & \no  & \no  & \yes & \no  \\
        CP-2.4 Sparse foreign key joins                   & \no  & \yes & \yes & \yes \\
        CP-3.2 Dimensional clustering                     & \no  & \yes & \no  & \no  \\
        CP-3.3 Scattered index access patterns            & \yes & \yes & \yes & \yes \\
        CP-5.3 Intra-query result reuse                   & \yes & \no  & \yes & \no  \\
        CP-7.2 Cardinality estimation of transitive paths & \yes & \no  & \yes & \yes \\
        CP-7.3 Execution of a transitive step             & \yes & \no  & \yes & \yes \\
        CP-7.5 Unweighted shortest paths                  & \yes & \no  & \yes & \yes \\
        CP-7.7 Composition of graph queries               & \yes & \yes & \yes & \yes \\
        CP-8.3 Ranking-style queries                      & \no  & \yes & \no  & \no  \\
        CP-8.4 Query composition                          & \yes & \yes & \yes & \yes \\
        CP-8.6 Handling paths                             & \yes & \no  & \yes & \yes \\
        \bottomrule
    \end{tabular}
    \caption{Relevance of the SNB choke points for the queries in the contest. Choke points not covered by the queries were omitted.
    \emph{Notation} --
        \yes:~relevant,
        \no:~not relevant.}
    \label{tab:choke-points}
\end{table}

The LDBC Social Network Benchmark~\cite{LdbcTechReport} uses \emph{choke points}~\cite{DBLP:conf/tpctc/BonczNE13,DBLP:journals/pvldb/DreselerBRU20} to characterize the difficulty of its queries.
These are challenging aspects of query optimization and evaluation, which have a significant impact on a system's performance when processing the queries. Choke points are divided into categories such as aggregation performance, join performance, data access locality, etc.

We have analyzed the queries of the contest in terms of the choke points provided in the latest SNB specification, version \textsf{0.4.0-SNAPSHOT}~\cite{LdbcTechReport}.
The choke points covered by the queries are shown in \autoref{tab:choke-points}.
The analysis confirms that the queries in the contest are complex: on average, a query has $35 / 4 = 8.75$ relevant choke points, making them quantifiably more complex than the queries of the Interactive workload, which have an average of 4.36 choke points/query, and the queries of the BI worload, which have 6.45.

\subsection{Contest setup}
\label{sec:contest-setup}

The contest defined a measurement scenario where the solutions-under-benchmark were given a set of queries and their input parameters.
Solutions competed on the total execution time, including the time of loading the data and evaluating the queries.
Deciding which preprocessing computations to perform (if any),
the order of queries to evaluate,
and how to parallelize the evaluation
was left for the solution to decide.

\begin{figure}[htb]
    \includegraphics[width=\linewidth]{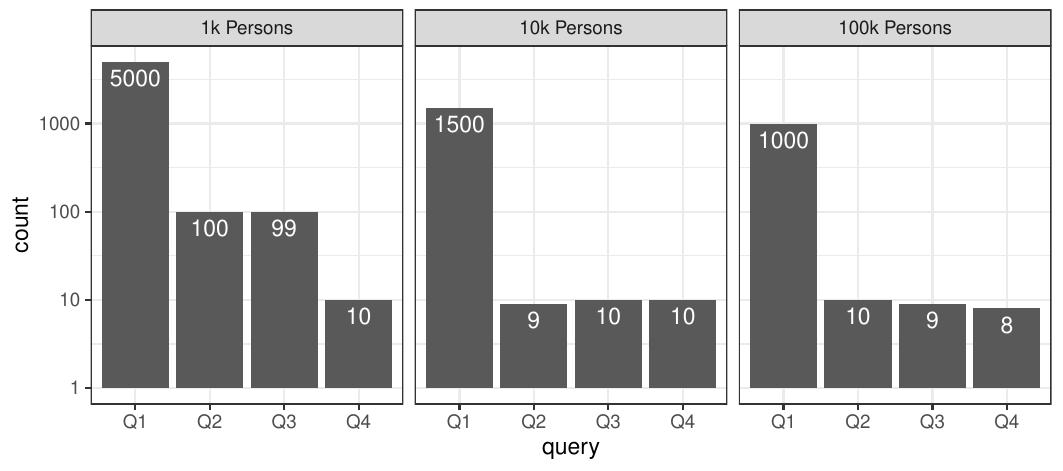}
    \caption{Number of queries used in the original contest for the 1k, 10k, and 100k \textsf{Person} data sets. Note that the $y$ axis is logarithmic.}
    \label{fig:query-statistics}
\end{figure}

The contest had an uneven mix of queries, with more than 95\% of the queries selected from Q1. The number of query instances per type for each data set is shown in \autoref{fig:query-statistics}.

\subsection{Parameter selection}

Due to the highly skewed distribution of the \textsf{Person}--\texttt{knows}--\textsf{Person} graph and the correlated nature of the data, using uniform random sampling will result in unpredictable query performance as demonstrated in the results of~\cite{HPEC2020_GraphBLAS_SIGMOD_Contest}.
If this is undesirable, we recommend using the \emph{parameter curation} technique defined in~\cite{DBLP:conf/tpctc/GubichevB14} and selecting representative parameters categories for each query.\footnote{The work on parameter curation was done as part of the LDBC project and it was published after the SIGMOD 2014 Contest. Therefore, parameter curation was presumably not used in the contest.} For example, for Q1, one might find that the representative categories are the Cartesian product of the following aspects:
\begin{itemize}
    \item \textbf{Subgraph computation:} no induced subgraph, induced subgraph with a low $x$ value, induced subgraph with a high $x$ value.
    \item \textbf{Reachability:} \textsf{Person} pairs that are unreachable from each other, reachable through few hops, reachable through many hops. 
\end{itemize}

\section{Influence}

\paragraph{Contest}
Overall, 33 teams have participated in the SIGMOD 2014 Programming Contest. The 5~top-ranking teams created posters and presentations detailing their solutions.
These were presented at SIGMOD 2014 in Snowbird, Utah (USA), where the awards ceremony also took place.

\paragraph{Followup works}
In the last 6 years, the contest has influenced a number of works in the graph processing space.
We provide a (non-exhaustive) list of papers and events that use this contest as their case study:

\begin{itemize}
    \item Team \emph{GenericPeople} from the Saint-Petersburg University (Russia) reported on their experience and discussed potential optimization ideas in paper~\cite{DBLP:conf/rcdl/ChernishevSSS14} and technical report~\cite{ChernishevReport}.

    \item The creators of two top-ranking solutions, \emph{AWFY} (TUM, Germany) and \emph{VIDA} (NYU, USA) co-authored a paper on efficient multi-source BFS~\cite{DBLP:journals/pvldb/ThenKCHPK0V14}.
          A continuation of this work from researchers at TUM is paper~\cite{DBLP:conf/edbt/KaufmannTK017} which presents a multi-threaded variant of the multi-source BFS algorithm.

    \item Members of the \emph{AWFY} team have co-authored papers on how to evaluate graph queries efficiently~\cite{DBLP:conf/btw/Kaufmann15} and on how to compute centrality values~\cite{DBLP:conf/btw/ThenGK017,DBLP:journals/dbsk/ThenGKN17}.

    \item In a loosely related line of work, members of the \emph{AWFY} team have co-authored a paper discussing the challenges of \emph{graph loading}, including node id relabelling, and compared potential approaches~\cite{DBLP:conf/grades/ThenKK016}.

    \item Queries similar to Q1 and Q3, \ie ones that compute shortest distances between \textsf{Persons} pairs or \textsf{Person} sets, are included in the LDBC Social Network Benchmark's \emph{Interactive}~\cite{DBLP:conf/sigmod/ErlingALCGPPB15,Angles2020} and \emph{Business Intelligence} workloads~\cite{DBLP:conf/grades/SzarnyasPAMPKEB18,LdbcTechReport}.

    \item The Grand Challenge of the DEBS 2016 (Distributed Event-Based Systems) conference defined a problem based on the data sets of the SIGMOD 2014 Programming Contest.%
          \footnote{\url{https://web.archive.org/web/20190131182518/https://debs.org/debs-2016-grand-challenge-social-networks/}}
          Its description states the following:

          \begin{quote}
              The data for the DEBS 2016 Grand Challenge is based on the dataset provided together with the LDBC Social Network Benchmark. DEBS 2016 Grand Challenge takes up the general scenario from the 2014 SIGMOD Programming Contest, however, in contrast to the SIGMOD contest, it explicitly focuses on processing streaming data and thus dynamic graphs.
          \end{quote}

    \item The 2018 Transformation Tool Contest, an annual contest held at the STAF (Software Technologies: Applications and Foundations) federation of conferences, presented the \emph{Social Media} benchmark~\cite{DBLP:conf/staf/Hinkel18} as its live case study which participants were required to solve during the conference.
          Similarly to the DEBS 2016 Grand Challenge, this case study defined two queries over a simplified social network schema which are evaluated on a continuously changing data set (thus favouring solutions that employ \emph{incremental view maintenance} techniques).

    \item In 2020, we have created a solution using the linear algebra-based concepts of the GraphBLAS C API and implemented in the SuiteSparse:GraphBLAS parallel library~\cite{DBLP:journals/toms/Davis19}.
          Our solution was published in paper~\cite{HPEC2020_GraphBLAS_SIGMOD_Contest} and is available online.\footnote{\url{https://github.com/ldbc/sigmod2014-contest-graphblas}}
\end{itemize}

\section{Online resources}

The source code, presentations, and posters of the top-5 teams are available on the contest's original website.\footnote{\url{http://www.cs.albany.edu/~sigmod14contest/leaders.html}}
To provide strong baseline implementations for evaluating our GraphBLAS solution presented in~\cite{HPEC2020_GraphBLAS_SIGMOD_Contest}, we have updated the solutions of two teams, \emph{AWFY} and \emph{blxlrsmb}, and made them available online.\footnote{\url{https://github.com/ftsrg/sigmod2014-pc-top-solutions}}
The updated solutions can be compiled with GCC version 9 and we made it possible to run each query implementation individually (compared to the contest's setup where multiple queries were expected to run in parallel).

\section*{Acknowledgements}

We would like to thank Jeong-Hyon Hwang for providing us the parameter files.
We would also like to thank Attila Nagy and Dávid Sándor for thorough discussions on the graph queries of the contest.

% biblatex
\printbibliography

@inproceedings{DBLP:conf/sigmod/ErlingALCGPPB15,
  xxauthor    = {Orri Erling and others},
  author    = {Orri Erling and
  Alex Averbuch and
  Josep{-}Lluis Larriba{-}Pey and
  Hassan Chafi and
  Andrey Gubichev and
  Arnau Prat{-}P{\'{e}}rez and
  Minh{-}Duc Pham and
  Peter A. Boncz},
  title     = {The {LDBC Social Network Benchmark}: {Interactive} Workload},
  booktitle = {SIGMOD},
  pages     = {619--630},
  year      = {2015},
  url       = {http://doi.acm.org/10.1145/2723372.2742786},
  doi       = {10.1145/2723372.2742786},
  bibsource = {dblp computer science bibliography, http://dblp.org}
}

@inproceedings{DBLP:conf/rcdl/ChernishevSSS14,
  author    = {George A. Chernishev and
               Vsevolod Sevostyanov and
               Kirill Smirnov and
               Ilya Shkuratov},
  title     = {On Several Social Network Analysis Problems},
  booktitle = {All-Russian Scientific Conference ``Digital libraries: Advanced Methods and Technologies, Digital Collections''},
  xxbooktitle = {Selected Papers of {XVI} All-Russian Scientific Conference "Digital
               libraries: Advanced Methods and Technologies, Digital Collections",
               Dubna, Russia, October 13-16, 2014},
  xxseries    = {{CEUR} Workshop Proceedings},
  xxvolume    = {1297},
  pages     = {234--242},
  xxpublisher = {CEUR-WS.org},
  year      = {2014},
  url       = {http://ceur-ws.org/Vol-1297/234-242_paper-34.pdf},
  bibsource = {dblp computer science bibliography, https://dblp.org}
}

@techreport{ChernishevReport,
  author      = {George A. Chernishev and
                 Vsevolod Sevostyanov and
                 Kirill Smirnov and
                 Ilya Shkuratov},
  title       = {On Several Social Network Analysis Problems: {A} Report},
  url         = {https://www.math.spbu.ru/user/chernishev/papers/sigmod2014contest-report.pdf},
  year        = {2014},
  institution = {Saint-Petersburg University, Russia},
}

@article{DBLP:journals/pvldb/ThenKCHPK0V14,
  author    = {Manuel Then and
               Moritz Kaufmann and
               Fernando Chirigati and
               Tuan{-}Anh Hoang{-}Vu and
               Kien Pham and
               Alfons Kemper and
               Thomas Neumann and
               Huy T. Vo},
  title     = {The More the Merrier: {E}fficient Multi-Source Graph Traversal},
  journal   = {{PVLDB}},
  volume    = {8},
  number    = {4},
  pages     = {449--460},
  year      = {2014},
  url       = {http://www.vldb.org/pvldb/vol8/p449-then.pdf},
  doi       = {10.14778/2735496.2735507},
  bibsource = {dblp computer science bibliography, https://dblp.org}
}

@inproceedings{DBLP:conf/edbt/KaufmannTK017,
  author    = {Moritz Kaufmann and
               Manuel Then and
               Alfons Kemper and
               Thomas Neumann},
  title     = {Parallel Array-Based Single- and Multi-Source Breadth First Searches
               on Large Dense Graphs},
  booktitle = {EDBT},
  xxpages     = {1--12},
  year      = {2017},
  url       = {https://doi.org/10.5441/002/edbt.2017.02},
  doi       = {10.5441/002/edbt.2017.02},
  bibsource = {dblp computer science bibliography, https://dblp.org}
}

@article{DBLP:journals/dbsk/ThenGKN17,
  author    = {Manuel Then and
               Stephan G{\"{u}}nnemann and
               Alfons Kemper and
               Thomas Neumann},
  title     = {Efficient Batched Distance, Closeness and Betweenness Centrality Computation
               in Unweighted and Weighted Graphs},
  journal   = {Datenbank-Spektrum},
  volume    = {17},
  number    = {2},
  pages     = {169--182},
  year      = {2017},
  url       = {https://doi.org/10.1007/s13222-017-0261-x},
  doi       = {10.1007/s13222-017-0261-x},
  bibsource = {dblp computer science bibliography, https://dblp.org}
}

@inproceedings{DBLP:conf/grades/ThenKK016,
  author    = {Manuel Then and
               Moritz Kaufmann and
               Alfons Kemper and
               Thomas Neumann},
  title     = {Evaluation of parallel graph loading techniques},
  booktitle = {GRADES at SIGMOD},
  publisher = {{ACM}},
  year      = {2016},
  url       = {https://doi.org/10.1145/2960414.2960418},
  doi       = {10.1145/2960414.2960418},
  bibsource = {dblp computer science bibliography, https://dblp.org}
}

@Article{Angles2020,
  author        = {Renzo Angles and J{\'{a}}nos Benjamin Antal and Alex Averbuch and Peter A. Boncz and Orri Erling and Andrey Gubichev and Vlad Haprian and Moritz Kaufmann and Josep{-}Llu{\'{\i}}s Larriba{-}Pey and Norbert Mart{\'{\i}}nez{-}Bazan and J{\'{o}}zsef Marton and Marcus Paradies and Minh{-}Duc Pham and Arnau Prat{-}P{\'{e}}rez and Mirko Spasic and Benjamin A. Steer and G{\'{a}}bor Sz{\'{a}}rnyas and Jack Waudby},
  journal       = {CoRR},
  title         = {The {LDBC} {S}ocial {N}etwork {B}enchmark},
  year          = {2020},
  volume        = {abs/2001.02299},
  archiveprefix = {arXiv},
  bibsource     = {dblp computer science bibliography, https://dblp.org},
  eprint        = {2001.02299},
  url           = {http://arxiv.org/abs/2001.02299},
}

@inproceedings{DBLP:conf/tpctc/PhamBE12,
  author    = {Minh{-}Duc Pham and
               Peter A. Boncz and
               Orri Erling},
  title     = {{S3G2}: {A} Scalable Structure-Correlated Social Graph Generator},
  booktitle = {TPCTC},
  xxseries    = {Lecture Notes in Computer Science},
  xxvolume    = {7755},
  pages     = {156--172},
  publisher = {Springer},
  year      = {2012},
  url       = {https://doi.org/10.1007/978-3-642-36727-4_11},
  doi       = {10.1007/978-3-642-36727-4_11},
  bibsource = {dblp computer science bibliography, https://dblp.org}
}

@article{DBLP:journals/csur/AnglesABHRV17,
  author    = {Renzo Angles and
               Marcelo Arenas and
               Pablo Barcel{\'{o}} and
               Aidan Hogan and
               Juan L. Reutter and
               Domagoj Vrgoc},
  title     = {Foundations of Modern Query Languages for Graph Databases},
  journal   = {{ACM} Comput. Surv.},
  volume    = {50},
  number    = {5},
  pages     = {68:1--68:40},
  year      = {2017},
  url       = {https://doi.org/10.1145/3104031},
  doi       = {10.1145/3104031},
  bibsource = {dblp computer science bibliography, https://dblp.org}
}

@inproceedings{DBLP:conf/btw/ThenGK017,
  author    = {Manuel Then and
               Stephan G{\"{u}}nnemann and
               Alfons Kemper and
               Thomas Neumann},
  title     = {Efficient Batched Distance and Centrality Computation in Unweighted
               and Weighted Graphs},
  booktitle = {Datenbanksysteme f{\"{u}}r Business, Technologie und Web (BTW)},
  series    = {{LNI}},
  volume    = {{P-265}},
  pages     = {247--266},
  publisher = {{GI}},
  year      = {2017},
  url       = {https://dl.gi.de/20.500.12116/632},
  bibsource = {dblp computer science bibliography, https://dblp.org}
}

@inproceedings{DBLP:conf/btw/Kaufmann15,
  author    = {Moritz Kaufmann and
               Tobias M{\"{u}}hlbauer and
               Manuel Then and
               Andrey Gubichev and
               Alfons Kemper and
               Thomas Neumann},
  title     = {Hochperformante {A}nalyse von {G}raph-{D}atenbanken},
  booktitle = {Datenbanksysteme f{\"{u}}r Business, Technologie und Web (BTW)},
  series    = {{LNI}},
  volume    = {{P-241}},
  pages     = {311--330},
  publisher = {{GI}},
  year      = {2015},
  url       = {https://dl.gi.de/20.500.12116/2414},
  bibsource = {dblp computer science bibliography, https://dblp.org}
}

@inproceedings{HPEC2020_GraphBLAS_SIGMOD_Contest,
	author = {M{\'a}rton Elekes and Attila Nagy and D{\'a}vid S{\'a}ndor and J{\'a}nos Benjamin Antal and Timothy A. Davis and G{\'a}bor Sz{\'a}rnyas},
	title = {A {GraphBLAS} solution to the {SIGMOD} 2014 {P}rogramming {C}ontest using multi-source {BFS}},
	booktitle = {High Performance Extreme Computing (HPEC)},
  publisher = {IEEE},
	year = {2020},
}

@inproceedings{DBLP:conf/grades/SzarnyasPAMPKEB18,
  author    = {G{\'{a}}bor Sz{\'{a}}rnyas and
               Arnau Prat{-}P{\'{e}}rez and
               Alex Averbuch and
               J{\'{o}}zsef Marton and
               Marcus Paradies and
               Moritz Kaufmann and
               Orri Erling and
               Peter A. Boncz and
               Vlad Haprian and
               J{\'{a}}nos Benjamin Antal},
  title     = {An early look at the {LDBC} {S}ocial {N}etwork {B}enchmark's {B}usiness {I}ntelligence workload},
  booktitle = {{GRADES-NDA} at {SIGMOD/PODS}},
  pages     = {9:1--9:11},
  publisher = {{ACM}},
  year      = {2018},
  url       = {http://doi.acm.org/10.1145/3210259.3210268},
  doi       = {10.1145/3210259.3210268},
  bibsource = {dblp computer science bibliography, https://dblp.org}
}

@inproceedings{DBLP:conf/staf/Hinkel18,
  author    = {Georg Hinkel},
  title     = {The {TTC} 2018 {S}ocial {M}edia Case},
  booktitle = {Transformation Tool Contest, co-located with the 2018 Software Technologies: Applications and Foundations (TTC@STAF)},
  series    = {{CEUR} Workshop Proceedings},
  volume    = {2310},
  pages     = {39--43},
  publisher = {CEUR-WS.org},
  year      = {2018},
  url       = {http://ceur-ws.org/Vol-2310/paper5.pdf},
  bibsource = {dblp computer science bibliography, https://dblp.org}
}

@article{DBLP:journals/talg/Chan12,
  author    = {Timothy M. Chan},
  title     = {All-pairs shortest paths for unweighted undirected graphs in \emph{o}(\emph{mn}) time},
  journal   = {{ACM} Trans. Algorithms},
  volume    = {8},
  number    = {4},
  pages     = {34:1--34:17},
  year      = {2012},
  url       = {https://doi.org/10.1145/2344422.2344424},
  doi       = {10.1145/2344422.2344424},
  bibsource = {dblp computer science bibliography, https://dblp.org}
}

@article{DBLP:journals/pvldb/MuhlbauerRSRK013,
  author    = {Tobias M{\"{u}}hlbauer and
               Wolf R{\"{o}}diger and
               Robert Seilbeck and
               Angelika Reiser and
               Alfons Kemper and
               Thomas Neumann},
  title     = {Instant Loading for Main Memory Databases},
  xxjournal   = {Proc. {VLDB} Endow.},
  journal   = {VLDB},
  volume    = {6},
  number    = {14},
  pages     = {1702--1713},
  year      = {2013},
  url       = {http://www.vldb.org/pvldb/vol6/p1702-muehlbauer.pdf},
  doi       = {10.14778/2556549.2556555},
  bibsource = {dblp computer science bibliography, https://dblp.org}
}

@article{DBLP:journals/corr/abs-1806-01799,
  author    = {Maciej Besta and
               Torsten Hoefler},
  title     = {Survey and Taxonomy of Lossless Graph Compression and Space-Efficient
               Graph Representations},
  journal   = {CoRR},
  volume    = {abs/1806.01799},
  year      = {2018},
  url       = {http://arxiv.org/abs/1806.01799},
  archivePrefix = {arXiv},
  eprint    = {1806.01799},
  bibsource = {dblp computer science bibliography, https://dblp.org},
}

@inproceedings{DBLP:conf/tpctc/BonczNE13,
  author    = {Peter A. Boncz and
               Thomas Neumann and
               Orri Erling},
  title     = {{TPC-H} Analyzed: {H}idden Messages and Lessons Learned from an Influential Benchmark},
  xxbooktitle = {Performance Characterization and Benchmarking - 5th {TPC} Technology
               Conference, {TPCTC} 2013, Trento, Italy, August 26, 2013, Revised
               Selected Papers},
  booktitle = {{TPCTC}},
  pages     = {61--76},
  year      = {2013},
  url       = {https://doi.org/10.1007/978-3-319-04936-6_5},
  doi       = {10.1007/978-3-319-04936-6_5},
  bibsource = {dblp computer science bibliography, https://dblp.org}
}

@article{DBLP:journals/pvldb/DreselerBRU20,
  author    = {Markus Dreseler and
               Martin Boissier and
               Tilmann Rabl and
               Matthias Uflacker},
  title     = {Quantifying {TPC-H} Choke Points and Their Optimizations},
  journal   = {Proc. {VLDB} Endow.},
  volume    = {13},
  number    = {8},
  pages     = {1206--1220},
  year      = {2020},
  url       = {http://www.vldb.org/pvldb/vol13/p1206-dreseler.pdf},
  bibsource = {dblp computer science bibliography, https://dblp.org}
}

@techreport{LdbcTechReport,
	author = {{LDBC Social Network Benchmark task force}},
	title = {The {LDBC Social Network Benchmark} (version 0.4.0-SNAPSHOT)},
	institution = {Linked Data Benchmark Council},
	year = {2020},
	url = {https://ldbc.github.io/ldbc_snb_docs/ldbc-snb-specification.pdf}
}

@inproceedings{DBLP:conf/sc/BeamerAP12,
  author    = {Scott Beamer and
               Krste Asanovic and
               David A. Patterson},
  title     = {Direction-optimizing breadth-first search},
  booktitle = {{SC}},
  publisher = {{IEEE/ACM}},
  year      = {2012},
  url       = {https://doi.org/10.1109/SC.2012.50},
  doi       = {10.1109/SC.2012.50},
  bibsource = {dblp computer science bibliography, https://dblp.org}
}

@inproceedings{DBLP:conf/icde/OlsenLH14,
  author    = {Paul W. Olsen and
               Alan G. Labouseur and
               Jeong{-}Hyon Hwang},
  title     = {Efficient top-k closeness centrality search},
  booktitle = {{ICDE}},
  pages     = {196--207},
  publisher = {{IEEE} Computer Society},
  year      = {2014},
  url       = {https://doi.org/10.1109/ICDE.2014.6816651},
  doi       = {10.1109/ICDE.2014.6816651},
  bibsource = {dblp computer science bibliography, https://dblp.org}
}

@article{DBLP:journals/jcss/Seidel95,
  author    = {Raimund Seidel},
  title     = {On the All-Pairs-Shortest-Path Problem in Unweighted Undirected Graphs},
  journal   = {J. Comput. Syst. Sci.},
  volume    = {51},
  number    = {3},
  pages     = {400--403},
  year      = {1995},
  url       = {https://doi.org/10.1006/jcss.1995.1078},
  doi       = {10.1006/jcss.1995.1078},
  bibsource = {dblp computer science bibliography, https://dblp.org}
}

@inproceedings{DBLP:conf/ipps/0001BB18,
  author    = {Michael Sutton and
               Tal Ben{-}Nun and
               Amnon Barak},
  title     = {Optimizing Parallel Graph Connectivity Computation via Subgraph Sampling},
  booktitle = {{IPDPS}},
  pages     = {12--21},
  publisher = {{IEEE} Computer Society},
  year      = {2018},
  url       = {https://doi.org/10.1109/IPDPS.2018.00012},
  doi       = {10.1109/IPDPS.2018.00012},
  bibsource = {dblp computer science bibliography, https://dblp.org}
}

@inproceedings{DBLP:conf/icalp/Tiskin01,
  author    = {Alexandre Tiskin},
  title     = {All-Pairs Shortest Paths Computation in the {BSP} Model},
  booktitle = {{ICALP}},
  series    = {Lecture Notes in Computer Science},
  volume    = {2076},
  pages     = {178--189},
  publisher = {Springer},
  year      = {2001},
  url       = {https://doi.org/10.1007/3-540-48224-5_15},
  doi       = {10.1007/3-540-48224-5_15},
  bibsource = {dblp computer science bibliography, https://dblp.org}
}

@article{DBLP:journals/corr/SolomonikH15,
  author    = {Edgar Solomonik and
               Torsten Hoefler},
  title     = {Sparse Tensor Algebra as a Parallel Programming Model},
  journal   = {CoRR},
  volume    = {abs/1512.00066},
  year      = {2015},
  url       = {http://arxiv.org/abs/1512.00066},
  archivePrefix = {arXiv},
  eprint    = {1512.00066},
  bibsource = {dblp computer science bibliography, https://dblp.org}
}

@inproceedings{DBLP:conf/ipps/SolomonikBD13,
  author    = {Edgar Solomonik and
               Aydin Bulu{\c{c}} and
               James Demmel},
  title     = {Minimizing Communication in All-Pairs Shortest Paths},
  booktitle = {{IPDPS}},
  pages     = {548--559},
  publisher = {{IEEE} Computer Society},
  year      = {2013},
  url       = {https://doi.org/10.1109/IPDPS.2013.111},
  doi       = {10.1109/IPDPS.2013.111},
  bibsource = {dblp computer science bibliography, https://dblp.org}
}

@article{DBLP:journals/toms/Davis19,
  author    = {Timothy A. Davis},
  title     = {Algorithm 1000: {SuiteSparse:GraphBLAS}: {G}raph Algorithms in the Language
               of Sparse Linear Algebra},
  journal   = {{ACM} Trans. Math. Softw.},
  volume    = {45},
  number    = {4},
  pages     = {44:1--44:25},
  year      = {2019},
  url       = {https://doi.org/10.1145/3322125},
  doi       = {10.1145/3322125},
  bibsource = {dblp computer science bibliography, https://dblp.org}
}

@article{DBLP:journals/jal/ShiloachV82,
  author    = {Yossi Shiloach and
               Uzi Vishkin},
  title     = {An $O$(log n) Parallel Connectivity Algorithm},
  journal   = {J. Algorithms},
  volume    = {3},
  number    = {1},
  pages     = {57--67},
  year      = {1982},
  url       = {https://doi.org/10.1016/0196-6774(82)90008-6},
  doi       = {10.1016/0196-6774(82)90008-6},
  bibsource = {dblp computer science bibliography, https://dblp.org}
}

@article{DBLP:journals/siamcomp/Tarjan72,
  author    = {Robert Endre Tarjan},
  title     = {Depth-First Search and Linear Graph Algorithms},
  journal   = {{SIAM} J. Comput.},
  volume    = {1},
  number    = {2},
  pages     = {146--160},
  year      = {1972},
  url       = {https://doi.org/10.1137/0201010},
  doi       = {10.1137/0201010},
  bibsource = {dblp computer science bibliography, https://dblp.org}
}

@article{DBLP:journals/sigmod/Tropashko05,
  author    = {Vadim Tropashko},
  title     = {Nested intervals tree encoding in {SQL}},
  journal   = {{SIGMOD} Rec.},
  volume    = {34},
  number    = {2},
  pages     = {47--52},
  year      = {2005},
  url       = {https://doi.org/10.1145/1083784.1083793},
  doi       = {10.1145/1083784.1083793},
  bibsource = {dblp computer science bibliography, https://dblp.org}
}

@inproceedings{DBLP:conf/tpctc/GubichevB14,
  author    = {Andrey Gubichev and
               Peter A. Boncz},
  title     = {Parameter Curation for Benchmark Queries},
  booktitle = {{TPCTC}},
  series    = {Lecture Notes in Computer Science},
  volume    = {8904},
  pages     = {113--129},
  publisher = {Springer},
  year      = {2014},
  url       = {https://doi.org/10.1007/978-3-319-15350-6_8},
  doi       = {10.1007/978-3-319-15350-6_8},
  bibsource = {dblp computer science bibliography, https://dblp.org}
}

@book{DBLP:books/daglib/0009092,
  author    = {Yousef Saad},
  title     = {Iterative methods for sparse linear systems},
  publisher = {{SIAM}},
  year      = {2003},
  url       = {https://doi.org/10.1137/1.9780898718003},
  doi       = {10.1137/1.9780898718003},
  isbn      = {978-0-89871-534-7},
  bibsource = {dblp computer science bibliography, https://dblp.org}
}

@misc{cwi:ldbc-sigmod-data-sets,
	author = {Márton Elekes and Gábor Szárnyas},
	title = {SIGMOD 2014 Programming Contest graphs},
	doi = {10.25606/SURF.7506be02023d2582},
	howpublished = {\url{https://hdl.handle.net/11112/dde63984-08bb-848c-4f00-bfdad71ed649}}
}

% bibtex
% \bibliographystyle{plainurl}
% \bibliography{ms}

\clearpage

\appendix

\section{Configuration of the LDBC Datagen}
\label{sec:datagen-params}

We list the configuration files used for generating the data sets used in the contest.
These configurations were passed in the form of the \texttt{params.ini} to the Datagen.

\begin{lstlisting}[caption=Configuration for the 100k \textsf{Person} data set.]
numtotalUser:100000
startYear:2013
numYears:1
serializerType:csv
\end{lstlisting}
\begin{lstlisting}[caption=Configuration for the 250k \textsf{Person} data set.]
numtotalUser:250000
startYear:2010
numYears:2
serializerType:csv
\end{lstlisting}
\begin{lstlisting}[caption=Configuration for the 500k \textsf{Person} data set.]
numtotalUser:500000
startYear:2010
numYears:2
serializerType:csv
\end{lstlisting}
\begin{lstlisting}[caption=Configuration for the 1M \textsf{Person} data set.]
numtotalUser:1000000
startYear:2010
numYears:2
serializerType:csv
\end{lstlisting}

\section{Sample queries and answers}
\label{sec:sample-queries-and-answers}

We present sample query parameters and expected outputs (answers) for the \texttt{1k} data set.

\paragraph{Note}
There is a one-to-one mapping between queries and answers. The answers files contain comments after the \texttt{\%} character. These are for debugging purposes. Files produced by solutions must not contain such comments.

\begin{lstlisting}[caption=\texttt{1k-sample-queries1.txt}]
query1(576, 400, -1)
query1(58, 402, 0)
query1(266, 106, -1)
query1(313, 523, -1)
query1(858, 587, 1)
query1(155, 355, -1)
query1(947, 771, -1)
query1(105, 608, 3)
query1(128, 751, -1)
query1(814, 641, 0)
\end{lstlisting}
\begin{lstlisting}[caption=\texttt{1k-sample-answers1.txt}]
3 % path 576-618-951-400 (other shortest paths may exist)
3 % path 58-935-808-402 (other shortest paths may exist)
3 % path 266-23-592-106 (other shortest paths may exist)
-1 % path none
4 % path 858-46-31-162-587 (other shortest paths may exist)
3 % path 155-21-0-355 (other shortest paths may exist)
2 % path 947-625-771 (other shortest paths may exist)
-1 % path none
3 % path 128-459-76-751 (other shortest paths may exist)
3 % path 814-109-557-641 (other shortest paths may exist)
\end{lstlisting}
\begin{lstlisting}[caption=\texttt{1k-sample-queries2.txt}]
query2(3, 1980-02-01)
query2(4, 1981-03-10)
query2(3, 1982-03-29)
query2(3, 1983-05-09)
query2(5, 1984-07-02)
query2(3, 1985-05-31)
query2(3, 1986-06-14)
query2(7, 1987-06-24)
query2(3, 1988-11-10)
query2(4, 1990-01-25)
\end{lstlisting}
\begin{lstlisting}[caption=\texttt{1k-sample-answers2.txt}]
Chiang_Kai-shek Augustine_of_Hippo Napoleon % component sizes 22 16 16
Chiang_Kai-shek Napoleon Mohandas_Karamchand_Gandhi Sukarno % component sizes 17 13 11 11
Chiang_Kai-shek Mohandas_Karamchand_Gandhi Napoleon % component sizes 13 11 10
Chiang_Kai-shek Mohandas_Karamchand_Gandhi Augustine_of_Hippo % component sizes 12 10 8
Chiang_Kai-shek Aristotle Mohandas_Karamchand_Gandhi Augustine_of_Hippo Fidel_Castro % component sizes 10 7 6 5 5
Chiang_Kai-shek Mohandas_Karamchand_Gandhi Joseph_Stalin % component sizes 6 6 5
Chiang_Kai-shek Mohandas_Karamchand_Gandhi Joseph_Stalin % component sizes 6 6 5
Chiang_Kai-shek Augustine_of_Hippo Genghis_Khan Haile_Selassie_I Karl_Marx Lyndon_B._Johnson Robert_John_\"Mutt\"_Lange % component sizes 4 3 3 3 3 3 3
Aristotle Ho_Chi_Minh Karl_Marx % component sizes 2 2 2
Arthur_Conan_Doyle Ashoka Barack_Obama Benito_Mussolini % component sizes 1 1 1 1
\end{lstlisting}
\begin{lstlisting}[caption=\texttt{1k-sample-queries3.txt}]
query3(3, 2, Asia)
query3(4, 3, Indonesia)
query3(3, 2, Egypt)
query3(3, 2, Italy)
query3(5, 4, Chengdu)
query3(3, 2, Peru)
query3(3, 2, Democratic_Republic_of_the_Congo)
query3(7, 6, Ankara)
query3(3, 2, Luoyang)
query3(4, 3, Taiwan)
\end{lstlisting}
\begin{lstlisting}[caption=\texttt{1k-sample-answers3.txt}]
361|812 174|280 280|812 % common interest counts 4 3 3
396|398 363|367 363|368 363|372 % common interest counts 2 1 1 1
110|116 106|110 106|112 % common interest counts 1 0 0
420|825 421|424 10|414 % common interest counts 1 1 0
590|650 590|658 590|614 590|629 590|638 % common interest counts 1 1 0 0 0
65|766 65|767 65|863 % common interest counts 0 0 0
99|100 99|101 99|102 % common interest counts 0 0 0
891|898 890|891 890|895 890|898 890|902 891|895 891|902 % common interest counts 1 0 0 0 0 0 0
565|625 653|726 565|653 % common interest counts 2 1 0
795|798 797|798 567|795 567|796 % common interest counts 1 1 0 0
\end{lstlisting}
\begin{lstlisting}[caption=\texttt{1k-sample-queries4.txt}]
query4(3, Bill_Clinton)
query4(4, Napoleon)
query4(3, Chiang_Kai-shek)
query4(3, Charles_Darwin)
query4(5, Ronald_Reagan)
query4(3, Aristotle)
query4(3, George_W._Bush)
query4(7, Tony_Blair)
query4(3, William_Shakespeare)
query4(4, Augustine_of_Hippo)
\end{lstlisting}
\begin{lstlisting}[caption=\texttt{1k-sample-answers4.txt}]
385 492 819 % centrality values 0.5290135396518375 0.5259615384615384 0.5249520153550864
722 530 366 316 % centrality values 0.5411255411255411 0.5405405405405406 0.5387931034482758 0.5382131324004306
592 565 625 % centrality values 0.5453460620525059 0.5421115065243179 0.5408284023668639
458 305 913 % centrality values 0.5415676959619953 0.5371024734982333 0.5345838218053928
953 294 23 100 405 % centrality values 0.5446859903381642 0.5394736842105263 0.5388291517323776 0.5375446960667462 0.5343601895734598
426 819 429 % centrality values 0.5451219512195121 0.5424757281553397 0.5366146458583433
323 101 541 % centrality values 0.553596806524207 0.5433196380862576 0.5413098243818695
465 647 366 722 194 135 336 % centrality values 0.535629340423861 0.5317101013475888 0.5310624641961301 0.530416402804164 0.5297719114277313 0.5259376153257211 0.5253039555482203
424 842 23 % centrality values 0.5316276893212858 0.5296265813313688 0.5256692220686188
385 562 659 323 % centrality values 0.5506329113924051 0.54375 0.54375 0.5291970802919708
\end{lstlisting}

\end{document}